\documentclass[twocolumn,showpacs,amsmath,amssymb]{revtex4}
\usepackage{dcolumn}

\usepackage{hyperref}
\usepackage{graphicx}
\usepackage{dcolumn}
\usepackage{bm}

\usepackage{slashed}

\def\+{\!+\!}
\def\-{\!-\!}
\def\={\!=\!}

\begin{document}

\title{Entanglement Entropy, Chemical Potential, Current Source, and Wilson Loop}

\author{Bom Soo Kim}
\affiliation{%
Physics Department, Loyola University Maryland, Baltimore, MD 21210 \\
Department of Physics and Astronomy, University of Kentucky, Lexington, KY 40506
}%

\date{\today}

\begin{abstract}
We construct and analytically compute entanglement and the R\'enyi entropies of Dirac fermions on a 2 dimensional torus in the presence of background chemical potential, current source and Wilson loop, by employing correlation functions of the electromagnetic vertex operators of $\mathbb{Z}_n$ orbifold theory. The entropies reveal numerous novelties. They probe the energy levels of compact fermions through the chemical potential, demonstrate interference phenomena with the current source being `beat frequency,' and experience topological phase transitions by dialing the Wilson loop, in the low temperature limit. In the large radius limit, they depend only on the Wilson loop via topological transitions, which are tightly linked to conformal dimension of the electromagnetic operators.
\end{abstract}


\maketitle

Entanglement is at the heart of quantum theories, encompassing quantum mechanics, quantum field theories, quantum gravity and quantum information science \cite{Einstein:1935rr}. Entanglement entropy \cite{Horodecki:2009zz,Calabrese:2004eu,Ryu:2006bv,Rangamani:2016dms,QunatumInformation} and its extension, the R\'enyi entropy \cite{Renyi}, can be used to measure quantum information encoded in a quantum state. The entropies have been useful to probe quantum critical phenomena \cite{Vidal:2002rm}, to classify topological states of matter that can not be distinguished by symmetries \cite{Kitaev:2005dm,Levin:2006zz}, and to prove the irreversibility of the renormalization group in 3 dimensional field theories that do not have other ways to do so \cite{Casini:2012ei}. Recently, the R\'enyi entropy has been measured in systems of interacting delocalized particles  using quantum interference of many body systems \cite{Experiment}. 

Direct computations of entanglement entropy on quantum field theories are known to be difficult. Nevertheless there have been progresses in 1+1 dimensions \cite{Bennett:1995tk,Holzhey:1994we,Calabrese:2009qy,Casini:2009sr}. Gauge fields and corresponding potentials are one of main tools for manipulating quantum fields. Time and space components of the 1+1 dimensional gauge potential are chemical potential $\mu$ and a source of current source $J$, respectively. We call the latter `current source' for simplicity. Entanglement entropy for quantum systems with finite chemical potential (or finite charge density) at zero temperature has been studied previously. The entropy was claimed to be independent of chemical potential for free fermions \cite{Ogawa:2011bz}\cite{Herzog:2013py} and for infinite quantum systems with an interval \cite{CardySlides:2016}. The R\'enyi entropy in an infinite system has been shown to have Wilson loop dependences (unfortunately, with singular entanglement entropy) \cite{Belin:2013uta}. Current source was also considered in a different setup \cite{Arias:2014ksa}. These unexpected results deserve systematic investigations from a general framework. Moreover, exact and analytic results on the entropies can shed lights on deeper understanding of the nature of quantum entanglement. 

In this letter, we provide a broader and complete picture of entanglement and the R\'enyi entropies for Dirac fermions on a 2 dimensional torus in the presence of chemical potential $\mu$, current source $J$, and/or topological Wilson loop $w$. We construct the most general entropy formula by employing correlation functions of the electromagnetic vertex operators in $\mathbb{Z}_n$ orbifold theories \cite{DiFrancesco:1997nk}\cite{Hori:2003ic}. The electric operator is identified as Wilson loop $w$, while the magnetic operator as the twist parameter of the orbifold $k$, where $ k\=-(n-1)/2, -(n-3)/2, \dots, (n-1)/2$. Conformal dimensions of the electromagnetic operators play crucial roles. We perform analytic and exact computations in the low temperature and large radius limits. We organize and summarize their salient features here. Details can be found elsewhere \cite{Kim:2017ghc}\cite{Kim:toAppear}.

First, in the zero temperature limit, the R\'enyi and entanglement entropies have non-trivial dependences on all the three, chemical potential $\mu$, current source $J$, and Wilson loops $w$. They pick up finite contributions whenever chemical potential coincides with one of the energy levels of the Dirac fermions. This  demonstrates their usefulness to probe the energy spectra of quantum systems. It is contrary to earlier results \cite{Ogawa:2011bz}\cite{Herzog:2013py}. 

The entropies are periodic in $J$, which plays the role of `beat frequency' when a modulus parameter $\alpha\= 2\pi\tau_1$ is dialed. The entropies of the periodic fermions are finite, while those of the anti-periodic fermions vanish.  This novel feature can be achieved by changing $\alpha$.

Wilson loops parameter $w$ also reveal the characteristics of the entropies in the zero temperature limit. Entanglement entropy experiences phase transitions at $w\=2p\-1$ for each integer $p$. As $w$ is increased, it shows an increasing tendency along with non-monotonicity within each transition range, $2p-1 \leq w < 2p+1$.     

Second, in the large radius limit, the entropies' dependences on chemical potential and current source vanish as fast as $\mathcal O \left(\ell_t/2\pi L\right)^2 $, where $\ell_t/2\pi L$ is size of sub-system over total system. This supports a recent claim that the entropies of an interval in infinite space are independent of $\mu$ \cite{CardySlides:2016}. We further generalize the claim for multiple intervals and in the presence of current source $J$. 

On the other hand, the entropies depend on the Wilson loops $w$, in the large radius limit, for the conformal dimension of vertex operators takes part in the spin structure independent entropies. This supports the claim \cite{Belin:2013uta} regarding the R\'enyi entropy and further generalizes to include entanglement entropy. 

Finally, we further compute the high temperature limit of the entropies and the mutual information in the presence of chemical potential, current source and Wilson loops. Their basic features can be found in \cite{Kim:2017ghc}\cite{Kim:toAppear}. The mutual information has the same functional dependences on the gauge fields and Wilson loops as those of the entropies. 
\\ \vspace{-0.1in}

{\bf Entanglement Entropy with $\mu, J$ and $w$:}
Entanglement entropy for an interval of length $\ell_t\=u\-v$ can be computed using the replica trick in terms of the $n$-copies of correlation functions of the vertex operators 
$$C_{w,k}\equiv \langle \sigma_{w,k} (u) \sigma_{-w,-k} (v) \rangle $$
in the $\mathbb{Z}_n$ orbifold theory for free fermions on a single 2 dimensional torus \cite{Casini:2005rm}\cite{Ogawa:2011bz}\cite{Herzog:2013py}. 
Where $\sigma_{\pm w \pm k}$ are the electromagnetic operators with an electric parameter $w$ representing Wilson loops and a magnetic parameter $k$ representing $k$-th twist of the $n$ replica copies. Their conformal dimensions play key roles and are given by 
\begin{align} \label{ConformalDimensionBody}
\Delta_{w,k}= \frac{1}{2} \alpha_{w,k}^2\;, \qquad \alpha_{w,k}= \frac{k}{n} + \frac{w}{2\pi} + l_k \;. 
\end{align} 
Where $l_k$ is an integer to ensure $-1/2 \leq \alpha_{w,k} <1/2$ when $w$ gets out of the range. We present further details in appendix \S \ref{sec:PartitionFunction}. 

Entanglement entropy is $n\! \to\! 1$ limit of the R\'enyi entropy  
\begin{align} \label{TopologicalEEFormula}
&S_n = \frac{1}{1- n}  \Big( \log Z[n] - n \log Z[1] \Big)  \;,
\end{align}
where $Z[n]= \prod_{k=-\frac{n-1}{2}}^{\frac{n-1}{2}} C_{w,k} $ and $ Z[1] = \lim_{n\to 1} Z[n] $. This normalization factor $Z[1]$ plays an important role in the presence of the Wilson loops. 
The correlation function factorizes as $C_{w,k}=C_{w,k}^0 \times C_{w,k}^{\mu J} $ 
\begin{align}
C_{w,k}^0 &=\bigg| \frac{2\pi \eta (\tau)^3 }{\vartheta [\substack{1/2 \\ 1/2 }](\frac{\ell_t}{L}|\tau)} \bigg|^{2\alpha_{w,k}^2} \;, \label{CorrelatorTwistedOp} \\
C_{w,k}^{\mu J} &=\bigg| \frac{\vartheta [\substack{1/2-a-J \\ b-1/2 }](\alpha_{w,k} \frac{\ell_t}{L} \+ \tau_1 J \+ i \tau_2 \mu|\tau)}{\vartheta [\substack{1/2-a-J \\ b-1/2 }](\tau_1 J + i \tau_2 \mu|\tau)} \bigg|^2\!\!. 
\label{CorrelatorTwistedOp2}
\end{align} 
Here the parameters $a$ and $b$ are twist boundary conditions along spatial and temporal cycles of a torus parametrized by $\tau\= \tau_1 \+ i \tau_2\= (\alpha + i \beta)/2\pi$, Dedekind function $\eta(\tau) \= q^{{1}/{24}} \prod_{n=1}^\infty (1 \- q^n)$ with $ q\= e^{2\pi i \tau}\!$, and Jacobi function $\vartheta [\substack{\alpha \\ \beta }] (z|\tau) = \sum_{n \in \mathbb Z} q^{(n + \alpha)^2/2} e^{2\pi i (z+\beta)(n + \alpha)}$. Following notations are more familiar:
$\vartheta_3(z|\tau) \= \vartheta \left[\substack{0 \\ 0} \right] (z|\tau)$, $\vartheta_2(z|\tau) \!\=\! \vartheta \left[\substack{0.5 \\ 0} \right]\! (z|\tau)$, $\vartheta_4(z|\tau) \!\=\! \vartheta \left[\substack{0 \\ 0.5} \right]\! (z|\tau)$, $\vartheta_1(z|\tau) \!\=\! \vartheta \left[\substack{0.5 \\ 0.5} \right] \!(z|\tau) $. 
$\vartheta_2$ is related to the periodic spatial and anti-periodic temporal circles, while $\vartheta_3$ to both anti-periodic circles. We focus on $\vartheta_2$ and $\vartheta_3$.

\begin{figure}[t]
	\begin{center}
		\includegraphics[width=.15\textwidth,height=.23\textheight,angle=-90]{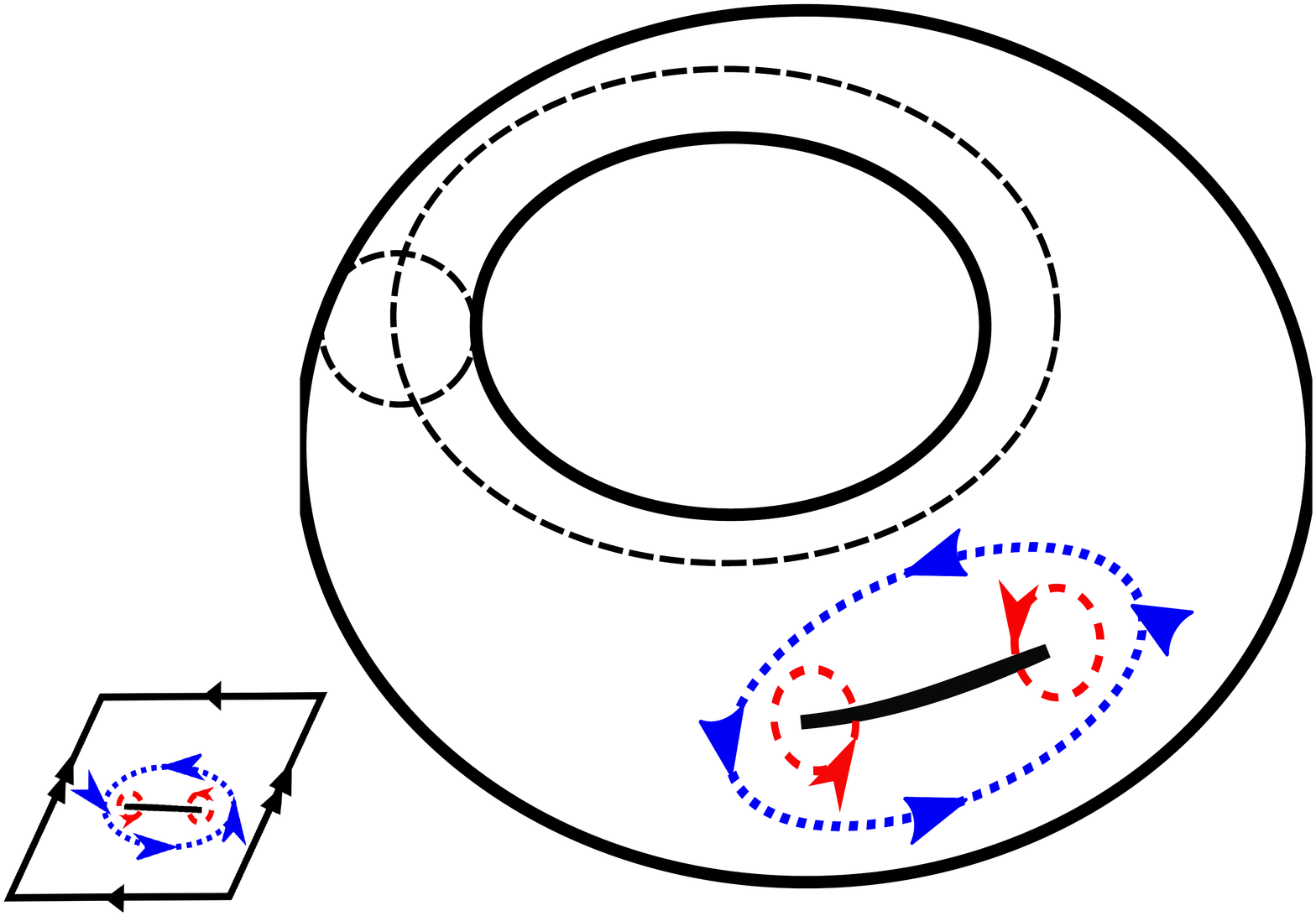} 
		\caption{\footnotesize\small Torus with a cut and the Wilson loop related to electric parameter $w$ (blue dotted line with arrows) along with the boundary conditions at both ends of the cut associated with the magnetic parameter $k$ (red dotted line with an arrow).  }
		\label{fig:TorusWithCuts11}
	\end{center}
\end{figure}

Equations \eqref{TopologicalEEFormula}-\eqref{CorrelatorTwistedOp2} are the most general formulas for the entropies in the presence of the background gauge fields and Wilson loops. Systematic and detailed treatments of the former in entropies have been done in \cite{Kim:2017ghc}, while the latter in \cite{Kim:toAppear}. See the figure \ref{fig:TorusWithCuts11}. Previously, the chemical potential dependence of the entropies was considered in \cite{Ogawa:2011bz}\cite{Herzog:2013py}\cite{CardySlides:2016}, while the Wilson loop dependence of the R\'enyi entropy in \cite{Belin:2013uta}. Generalization with multiple intervals is straightforward by using $\ell_t \= \sum_{a} (u_a \- v_a)$ when all the $w$'s are the same. (See {\it e.g.} \cite{Herzog:2013py} for $\mu\neq 0$.) 

The entropies decompose as $S_{w,n}=S_{w,n}^0 \+ S_{w,n}^{\mu J} $. $C_{w,k}^0$ is independent of $\mu$ and $J$, and so is $S_{w,n}^0$ \cite{Ogawa:2011bz}\cite{Herzog:2013py}, while they depend on $w$ explicitly through $\alpha_{w,k}$.  The rest of the letter investigates novel and interesting features of  \eqref{TopologicalEEFormula}-\eqref{CorrelatorTwistedOp2} focusing on various limits in the presence of the gauge fields and Wilson loops.  
\\ \vspace{-0.1in}

{\bf Entanglement entropy with chemical potential:} 
We first consider the entropies with a non-zero chemical potential $\mu$ (with $\tau_1 \=J\= w\= 0$) {\it at the zero temperature limit $\beta\to \infty$.} It has been argued that entanglement entropy at zero temperature is independent of $\mu$ \cite{Ogawa:2011bz}\cite{Herzog:2013py}. 

Here we take a more refined limit, $\beta \to \infty$, $\mu-N/2 \to 0$ keeping $\beta (\mu-N/2) \to const.$ for integer $N$ to see a non-trivial chemical potential dependence even in the zero temperature limit. For anti-periodic fermions ($a\=1/2$), the R\'enyi entropy $S^{\mu}_n$ is 
\begin{align} \label{SAmu3}
S^\mu_n &\= \frac{1}{1\-n} \Big[ \sum_{k=-\frac{n-1}{2}}^{\frac{n-1}{2}} \!\!\!\! \log \Big|\frac{\vartheta_3 (\frac{k}{n} \frac{\ell_t}{2\pi L}\+\frac{i\beta\mu}{2\pi} |\frac{i\beta}{2\pi} )}{\vartheta_3 (\frac{i\beta\mu}{2\pi} |\frac{i\beta}{2\pi})} \Big|^2 ~\Big] \;. 
\end{align}  
We use the product representation of Jacobi theta function
$
\vartheta_3 (z|\tau) = \prod_{m=1}^\infty (1 \- q^m)(1 \+ y q^{m-\frac{1}{2}})(1 \+ y^{\-1} q^{m-\frac{1}{2}}),$ where $q\=e^{2\pi i \tau}, y\=e^{2\pi i z}
$, 
to get \cite{Kim:2017ghc}\cite{Nishioka:2009zj}
\begin{align}\label{SAmuNSRenyi}
S^\mu_n 
&\=\frac{2}{n\-1}\! \sum_{l=1}^{\infty} \frac{(-1)^{l-1}}{l} 
\frac{\cosh (l\beta\mu)}{\sinh ( \frac{l\beta}{2} )} \! \left[n\-  \frac{\sin \left(\! \frac{l \ell_t}{2L}\right)}{\sin \left(\! \frac{1}{n} \frac{l\ell_t}{2L}\right)} \! \right]. 
\end{align} 
Detailed computations are given in \eqref{app2} in appendix \S \ref{sec:AntiPeriodicZeroEE}.  
The limit $n\to 1$ gives entanglement entropy. 
\begin{align} \label{SAmuNS}
S^\mu_{n\to 1} 
&\=2\sum_{l=1}^{\infty} \frac{(-1)^{l-1}}{l} 
\frac{\cosh (l\beta\mu)}{\sinh ( \frac{l\beta}{2} )} F(\ell_t,\! L) \;, 
\end{align} 
where $F(\ell_t,\! L) \= \left[1\- \frac{l\ell_t}{2L} \cot \left(\frac{l\ell_t}{2L} \right) \right] $.
These results are valid for $e^{-\beta\mu-\beta/2} \!<\! 1$ and $e^{\beta\mu-\beta/2} \!<\! 1 $, and thus for $-1/2 \!<\! \mu \!<\! 1/2$. The entropies vanish at zero temperature. Nevertheless, this conclusion is not valid for $\mu = \pm 1/2$. Let us carefully look into them. 

For this purpose, we consider $\beta(\mu-1/2) \=M \= const.$ in the limit $\beta \!\to\! \infty$ and $\mu \to 1/2$.  
A modified expansion $\vartheta_3 (z|\tau) \= (1 \+ y^{-1} q^{\frac{1}{2}}) \prod_{m=1}^\infty (1 \- q^m)(1 \+ y q^{m\-\frac{1}{2}})(1 \+ y^{-1} q^{m+\frac{1}{2}}) $ is useful. The factor $(1 \+ y^{-1} q^{\frac{1}{2}}) 
 = 1\+ e^M$ is finite. While entanglement entropy exists for general $M$, we consider $M<1$ here for simplicity. 
Then $S^\mu_{n\to 1}$ is
\begin{align}
S^\mu_{n\to 1} = 2\!\sum_{l=1}^{\infty} \frac{(-1)^{l-1}}{l} \!
\bigg[e^{-Ml}\!\+ \frac{e^{-\frac{l\beta}{2}}}{\sinh(\frac{l\beta}{2})}  \bigg]  F(\ell_t,\! L). 
\end{align} 
Here $\mu \!\to\! 1/2$ and $|y_{1,2} q^{1/2}| \!<\! 1$ and $|y_{1,2}^{-1} q^{3/2}| \!<\! 1$ are used. The result is valid for $-1/2< \mu <3/2$.
The first term in the square parenthesis is non-zero in the zero temperature limit. 
More generally, there are non-zero contributions when $\beta \left(\mu - (2N+1)/2 \right) = const.$ for $\beta \!\to\! \infty$ and $\mu \!\to\! (2N+1)/2$ with an integer $N$, which is identified as the energy levels of the particles in a compactified circle with an anti-periodic boundary condition. 

Similarly, entanglement entropy for Dirac fermions with periodic boundary conditions ($a\=0$) picks up finite values when $\mu\! \to\! N$ at zero temperature. 
Combining them together, entanglement entropy reveals non-trivial contributions depending on the chemical potential when 
\begin{align}
\beta \left(\mu - N/2 \right) = const. \;, 
\end{align}
for $\beta \!\to\! \infty$ and $\mu \!\to\! N/2$, which is identified as the energy levels of the particles in a compactified circle. We expect this happens generically, providing a useful way to probe the energy levels of a given system.

We move on to discuss the entropies on $\mu$ dependence in {\it the large radius limit}. 
It has been argued that entanglement entropy is independent of $\mu$ for a single interval in an infinitely long space \cite{CardySlides:2016}. We support and generalize the claim by taking an infinite space limit $\ell_t/L \to 0 $ or a limit of a small system size. Our results are also valid for multiple intervals.  

The R\'enyi entropy $S^\mu_n $  \eqref{SAmu3} for the fermions with anti-periodic boundary conditions in both time and space circles is given in \eqref{app4} in appendix \S \ref{sec:AntiPeriodicLargeEE}.   
\begin{align} \label{EELargeSpaceLimitNSSectorRenyi}
\frac{(n\+1) \ell_t^2}{12n L^2}\! \sum_{m=1}^{\infty}\!\! \frac{ 1 \+ \cosh (\beta\mu) \cosh ([m\-\frac{1}{2}] \beta )}{(\cosh ([m\-\frac{1}{2}] \beta ) \+ \cosh (\beta \mu) )^2 }  \+  \cdots,  
\end{align} 
where $\ell_t/L \!\ll\! 1 $. Taking $n\! \to \! 1$ provides entanglement entropy. The entropies vanish at least as $ \ell_t^2/L^2 $ as the size approaches infinity. The periodic sector is similar. 

This result is valid for more general cases, including all the spin structure dependent entropies in the presence of chemical potential, current source and/or the Wilson loops, as explained further in \eqref{app5}. Nevertheless, the large radius limit depends on the Wilson loops parameter $w$ due to  $\alpha_{w,k}$ in \eqref{CorrelatorTwistedOp} as we see below. 
\\ \vspace{-0.1in}

{\bf Entanglement entropy with current source:} 
In \cite{Kim:2017ghc} and \S \ref{sec:PartitionFunction}, we have developed the partition function for the 2 dimensional Dirac fermion in the presence of background gauge fields $ \tilde A = \tilde \mu dt + \tilde J ds$. The following twist boundary conditions without the background fields 
\begin{align} \label{TwistedeBCBody}
\psi_\pm (t,s) &= e^{-2\pi i a} \psi_\pm (t,s+2\pi) \nonumber \\
&= e^{-2\pi i b} \psi_\pm (t+2\pi \tau_2, s+2\pi \tau_1)  
\end{align}
have an equivalent description. The correspondence can be summarized as $\tilde A \= a ds \+ \frac{b -a\tau_1}{\tau_2} dt$. Thus, 
\begin{align}
a = \tilde J \;,  \qquad b = \tau_1 \tilde J + i \tau_2 \tilde \mu \;.
\end{align}

We add current source $J$ to the boundary condition \eqref{TwistedeBCBody}. The mode expansion of the fermions is modified. 
\begin{align}
\psi_- =  \sum_{\tilde r \in \mathbb Z + a + J} \psi_r (t) e^{i \tilde r s} \;.
\end{align}
current source changes the periodicity of compact fermions and thus produces distinctive physical effects on  the entropies. For simplicity, let us fix $\tau_1\= 0$, $a\= 0$ and $\mu \= w \= 0$. We increase the current source from $J=0$ to $J=1/2$ in the zero temperature limit. Then the boundary condition in the spatial circle changes from periodic to anti-periodic. Explicitly, entanglement entropy goes 
\[ S_{n\to 1} \= \!\left\{ \begin{array}{ll}
\! \sum_{l=1}^{\infty} \frac{2(\-1)^{l\-1}}{l \sinh ( {l\beta}/{2} )} F(\ell_t,\! L) \;,  &  J\=0 \;, \\  & \vspace{-0.05in} \\
\! \sum_{l=1}^{\infty} \frac{2(\-1)^{l\-1}}{l \tanh ( {l\beta}/{2} )} F(\ell_t,\! L) \;,  &  J\=\frac{1}{2} \;. \end{array} \right. \] 
Increasing the current source $J$ brings visible effects in entanglement entropy, from a zero value to a non-zero value. 

\begin{figure}[t]
	\begin{center}
		\includegraphics[width=.23\textwidth]{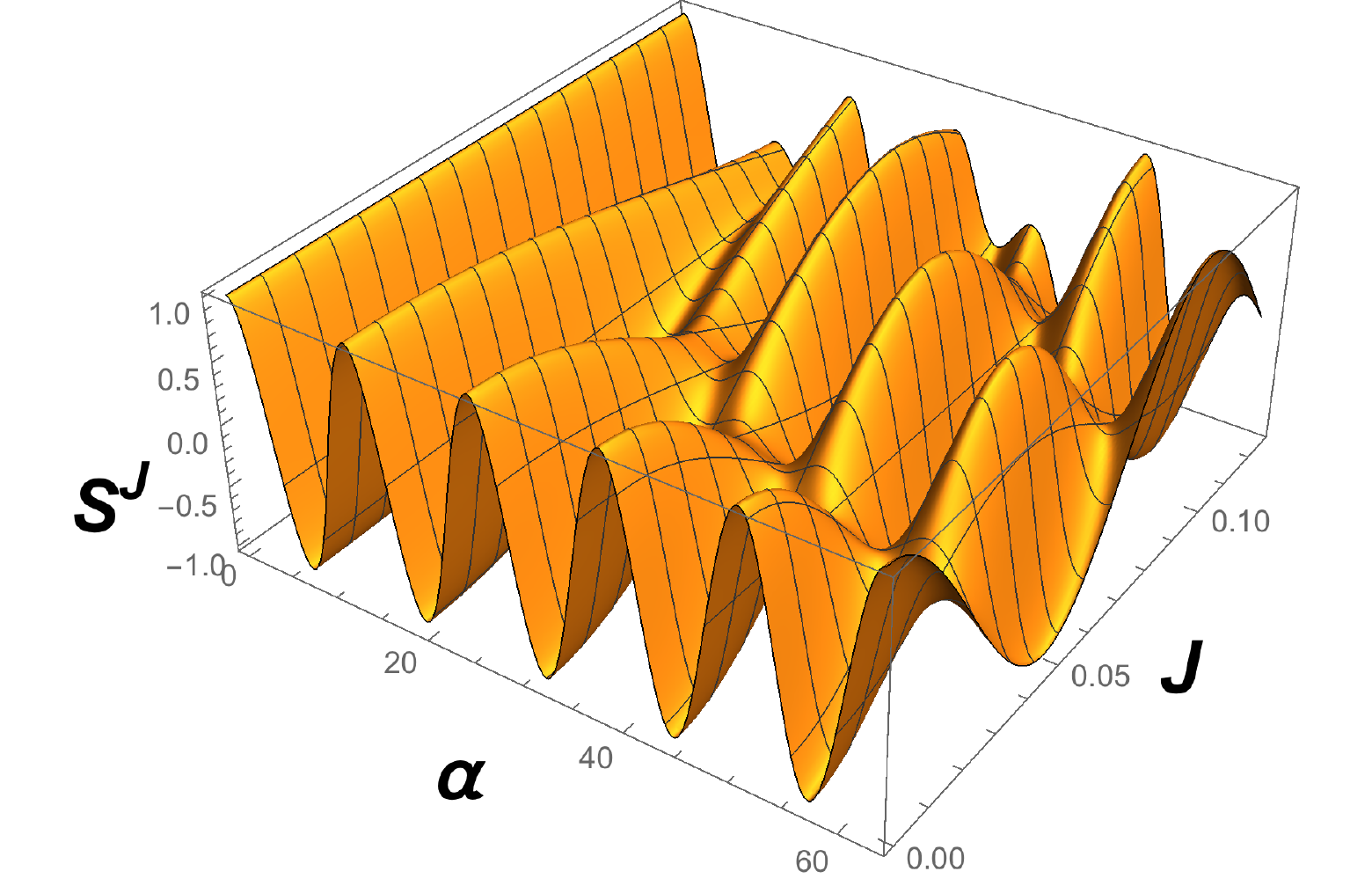} 
		\includegraphics[width=.23\textwidth]{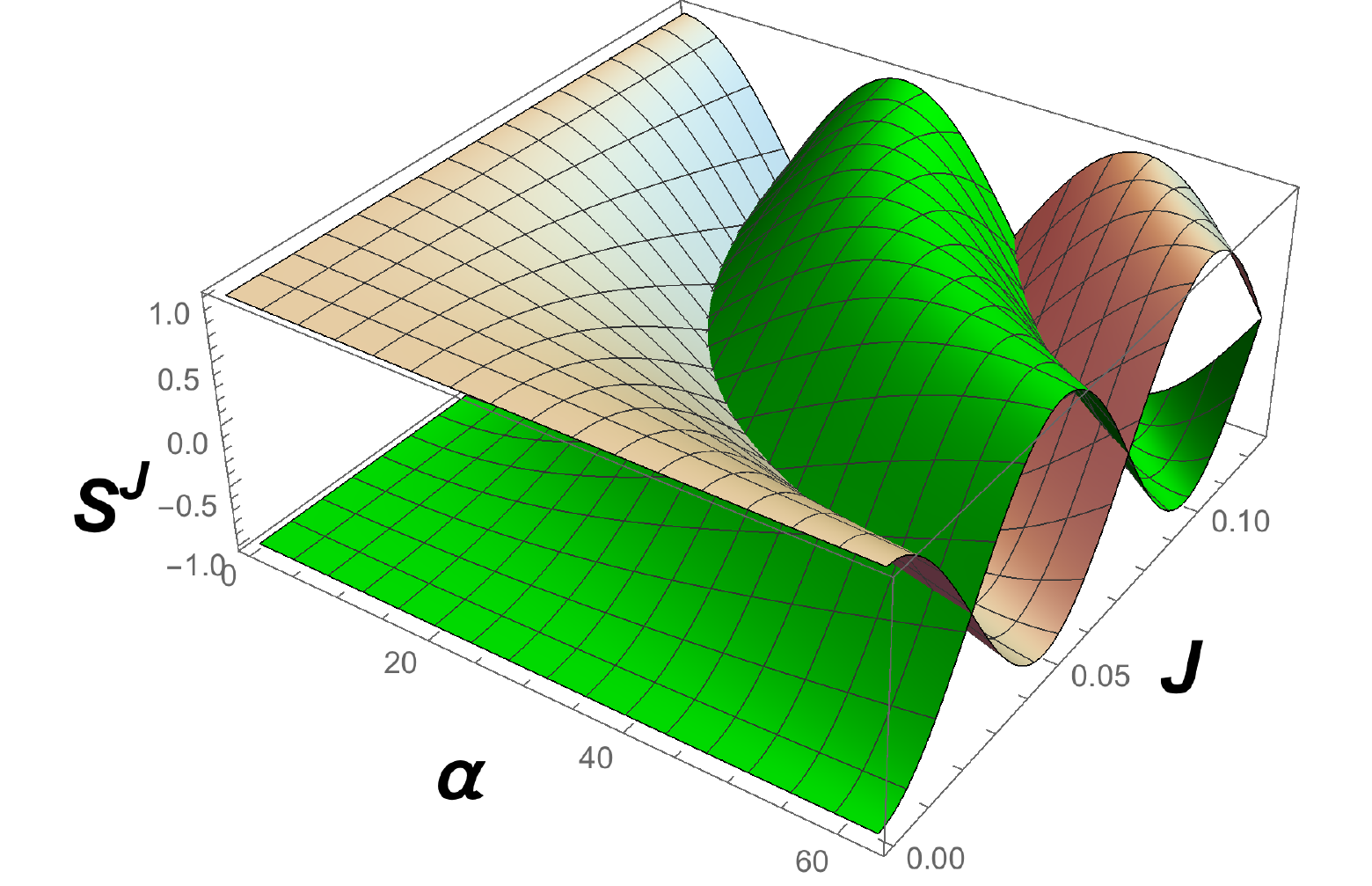} 
		\caption{\footnotesize\small Left: plot for the function given in \eqref{BeatFormula} for the ranges $0 \! \leq \alpha \!< 20\pi$ and $0 \!\leq J \!< 1/8$. Right: two enveloping beat functions for the left plot. Dialing $\alpha$ twists the torus.}
		\label{fig:BeatFrequency}
	\end{center}
\end{figure}

Motivated, we compute entanglement entropy with $J$ and $\alpha\= 2\pi \tau_1$ keeping $\mu\= w\= 0$ in the {\it zero temperature limit.} First, we consider {\it anti-periodic fermion}, $a \+ J\= 1/2$. 
\begin{align} \label{SAmuNSJ}
S^J_{1} 
&\= 4\!\! \sum_{l,m=1}^{\infty} \!\!\! \frac{(-1)^{l\-1} \cos (\alpha J l)}{l e^{(m-1/2) \beta l}}  \cos ([m\-\frac{1}{2}]\alpha l) F(\ell_t,\! L). 
\end{align} 
Entanglement entropy is a periodic function of $J$ and $\alpha$, which is different from the dependence on $\mu$. Interestingly, it depends on $\alpha$ in two different ways. The term with $l\=m\=1$ provides a dominant contribution 
\begin{align} \label{BeatFormula}
S^J_{1}(l\=m\=1)  \propto 4  e^{-\frac{\beta}{2}} \cos \left(\frac{\alpha}{2} \right)
\cos (\alpha J) \;. 
\end{align} 
When we dial the parameter $\alpha$ (twisting the spatial circle of torus) for a fixed $J$, the product of two cosine functions produce an `interference pattern.' For $J < 1/2$, one observes an interference with a `beat frequency' $ J/\pi $. See figure \ref{fig:BeatFrequency}. 
 
We also consider {\it periodic fermions,} $a+J\=0$. Similarly, 
\begin{align}\label{SAmuRJ22}
S^J
&\=  2 \sum_{l=1}^{\infty} \frac{(-1)^{l-1} \cos (\alpha J l)}{l} \Big[ 1 + \sum_{m=1}^{\infty} 2\frac{\cos (m\alpha l)}{e^{m \beta l}} \Big]  
\nonumber \\
&\qquad\qquad \times F(\ell_t,L) \;. 
\end{align} 
Entanglement entropy has a finite contribution depending on the current source $J$ at $\beta \!\to\! \infty$. This is drastically different from the results of anti-periodic fermions given in \eqref{SAmuNSJ}. As one dials the current source $J$ for a fixed $a$, the entanglement entropy changes from \eqref{SAmuNSJ} (that vanishes at $T=0$) to \eqref{SAmuRJ22}, which has a non-zero contribution. This effect is expected to be present in general quantum systems along with the interference and beat frequency.    

Entanglement entropy depends on current source only through $\alpha J$ and becomes negative for certain values of the combination. Total entropy is positive if one takes into account another contributions. 
\\ \vspace{-0.1in}

{\bf Entanglement entropy with Wilson loops:} 
The R\'enyi entropy in {\it the infinite space} with Wilson loops $w$ was considered in \cite{Belin:2013uta}, which shows that it depends on $w$ and the corresponding phase transitions. However, entanglement entropy was singular \cite{Belin:2013uta}. 

Here we generalize and carefully compute the Wilson loops $w$ dependence of the entropies in a finite space  \eqref{TopologicalEEFormula}-\eqref{CorrelatorTwistedOp2} to find the smooth limit for entanglement entropy by employing a different normalization $Z[1]$. We also show the entropies depend on $w$ in the large radius limit (infinite space) as well as low temperature limits. Details can be found in \cite{Kim:toAppear}.

The spin structure independent R\'enyi entropy $S_{w,n}^0$ has the form \eqref{TopologicalEEFormula}-\eqref{CorrelatorTwistedOp}. We start with 
\begin{align} \label{SpinIndRW1}
\log Z_w^0[n] = \bigg[ \sum_{k=-\frac{n-1}{2}}^{\frac{n-1}{2}} \alpha_{w,k}^2 \bigg] \log \Big|\frac{2\pi \eta (\tau)^3 }{\vartheta _1(\frac{\ell_t}{2\pi L}|\tau)} \Big|^2 \;, 
\end{align}
where $\alpha_{w,k}=\frac{k}{n} \+ \frac{w}{2\pi} \+ l_k$ is given in \eqref{ConformalDimensionBody}. As mentioned, the integer $l_k$ depends on $w$ such that $|\alpha_{w,k}| \leq 1/2$. As $w$ increases, some of $l_k$'s become non-zero, developing topological phase transitions. For $w \! < \! \pi/n$, all the $l_k$'s vanish.  
$S_{w,n}^0$ is simple to compute and independent of $w$.

\begin{figure}[t]
	\begin{center}
		\includegraphics[width=.23\textwidth]{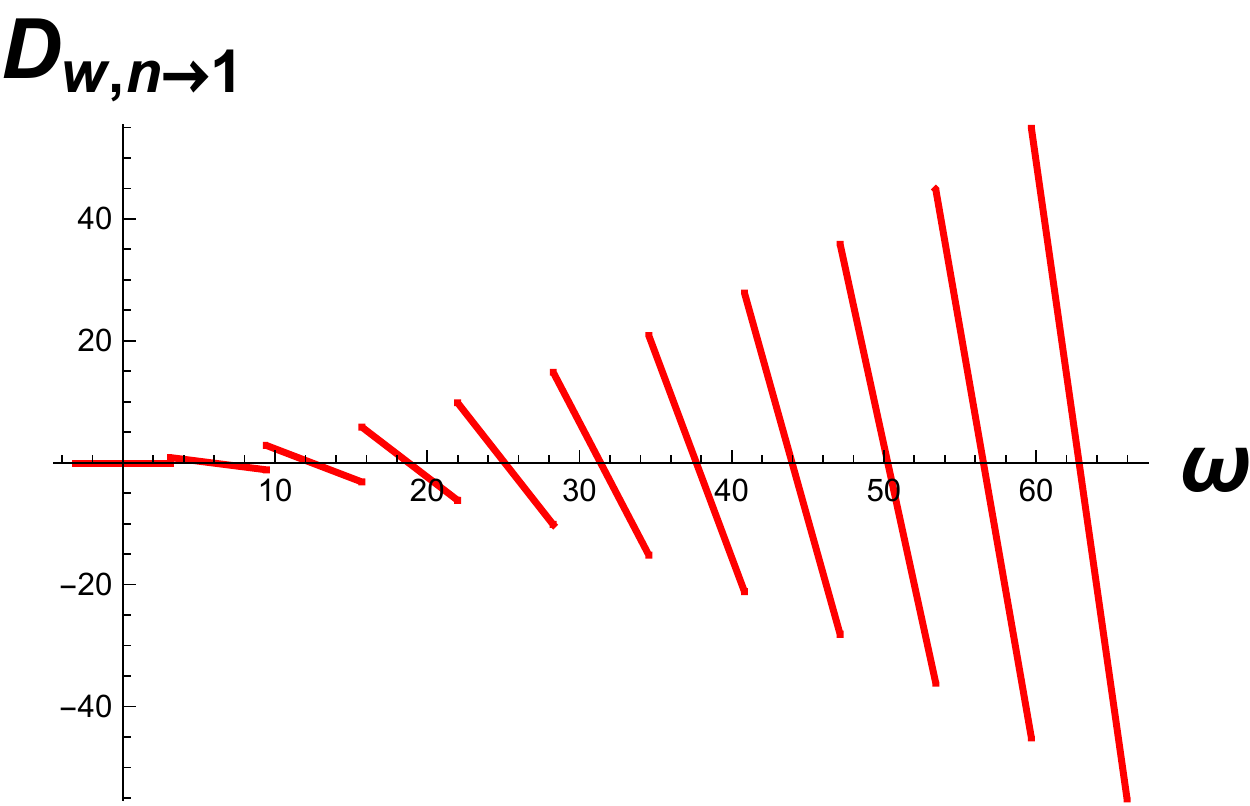} 
		\includegraphics[width=.23\textwidth]{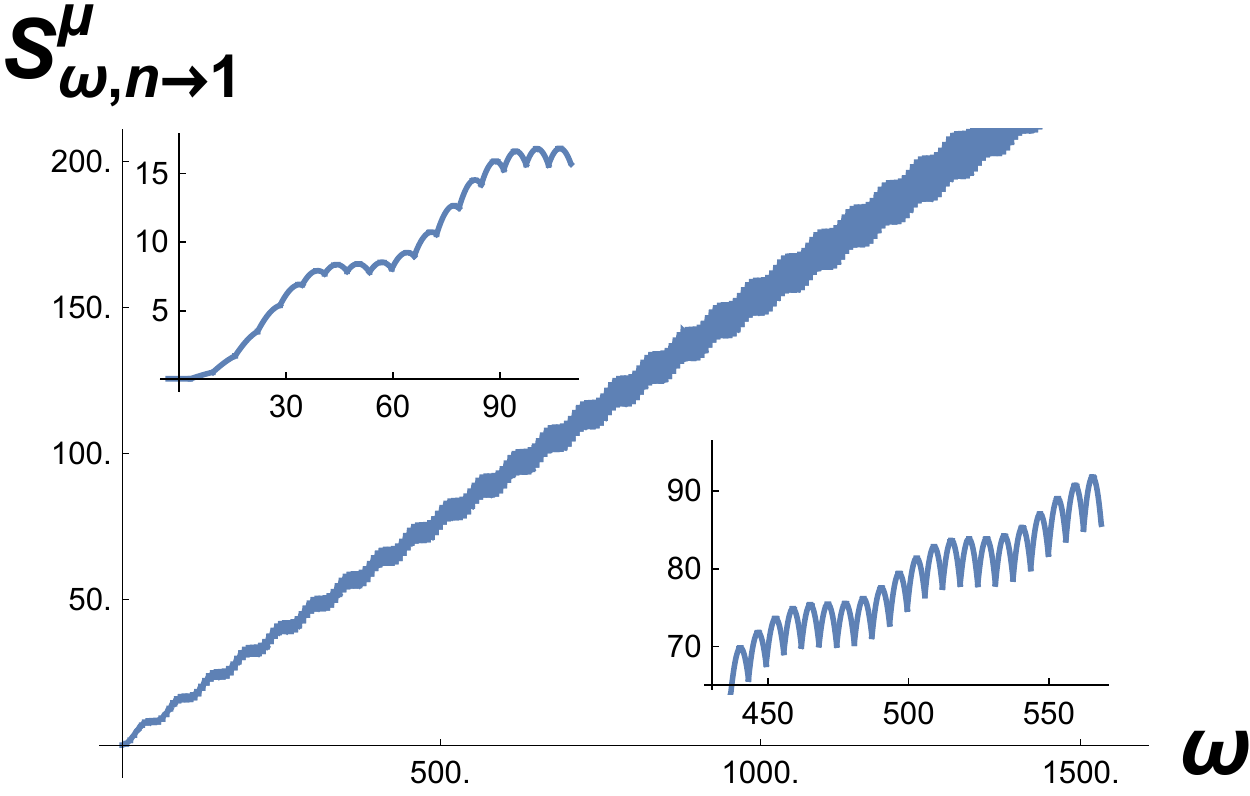} 
		\caption{\footnotesize\small Spin structure independent $S^0_{w,1}$ (\eqref{EEIndSpinSP=1}, left) and dependent $S^\mu_{w,1}$ (\eqref{LowTEEW33} with $l\!=\! 1, \ell_t/L\!=\!0.75$, right) entropies as a function of $w$. There are clear topological transitions. }
		\label{fig:WilsonLoops}
	\end{center}
\end{figure}

For $\pi/n  \!<\! w \! < \! 3\pi/n $, one of the $l_k$'s become non-zero, $l_{(n-1)/2} \=-1$. Then $\sum_{k} \alpha_{w,k}^2 \= \frac{n^2-1}{12 n} - \frac{w}{\pi} + \frac{1}{n} + n (  \frac{w}{2\pi} )^2$. Here we observe that $Z[1]$ is not clearly determined by any means. In particular, it can be different from $ Z[1]\= (\frac{w}{2\pi})^2$, which is for $w \! < \! \frac{\pi}{n}$ \cite{Belin:2013uta}. One important usage of the R\'enyi entropy is to compute entanglement entropy. To have a smooth entanglement limit, we employ $ Z[1] \= \lim_{n\to 1} Z[n] \=(  \frac{w}{2\pi} )^2 \- \frac{w}{\pi} \+ 1$. Then, 
\begin{align} \label{EEIndSpinSP=1}
S_{w,n}^{0} \=  D_{w,n} \log \Big|\frac{2\pi \eta (\tau)^3 }{\vartheta_1 (\frac{\ell_t}{2\pi L}|\tau)} \Big|^2  \;,  
\end{align}
where $D_{w,n}\= -\left(\frac{n+1}{12 n} + \frac{w}{\pi} - \frac{n+1}{n}\right) $.
This R\'enyi entropy has a smooth $n \!\to\! 1$ limit. This can be generalized for $ (2p-1)\pi/n \leq w < (2p+1) \pi/n$ with $p$-th topological transition. The entropies are given by \eqref{SpinIndRW1} with 
$ D_{w,n}\= - \Big( \frac{n+1}{12 n} + \frac{p(p+1)}{2} \big(\frac{w}{\pi}+1 \big) - \frac{2n+1}{n} \frac{p(p+1)(2p+1)}{6} \Big)$. This $D_{w,1}$ is depicted in the left figure \ref{fig:WilsonLoops}.  
We note that this spin structure independent entropies $S_{w,n}^{0}$ given in \eqref{EEIndSpinSP=1} has non-trivial Wilson loop dependences through $ D_{w,n}$ both in the large radius and low temperature limits. The latter case is due to the sine factor in $\vartheta_1$.

Spin structure dependent entropies $S_{w,n}^{\mu J}$ also show interesting $w$ dependences. Let us set $\alpha\=\mu\=J\=0$. We show their properties for the anti-periodic fermions $a\=b\=1/2$ in the low temperature limit. Computation is similar to that of \eqref{SAmu3} with $k/n \to \alpha_{w,k} $ and $\mu=0$. There are topological transitions depending on $w$. For general $p$-transitions, entanglement entropy goes \cite{Kim:toAppear}  
\begin{align} \label{LowTEEW33}
&S_{w,1}
\=\! \sum_{l=1}^{\infty} \frac{2(-1)^{l-1}}{l\sinh (\frac{l\beta }{2})}  
\Big[ \cos ( \frac{\ell_t l w}{2\pi L}) F(\ell_t,\! L) \bigg.  \\
&\+  p \cos ( \frac{\ell_t l (w \- 2p\pi)}{2\pi L } ) \- \cos ( \frac{\ell_t l (w \+ (1\- p)\pi)}{2\pi L } ) \frac{\sin (\frac{\ell_t l p }{2L } )}{\sin (\frac{\ell_t l }{2L } )}  
\Big]. \nonumber 
\end{align}
Note the term that is linear in $p$, which increases stepwise and drives the entropy to increase as $w$ increases. See the right figure \ref{fig:WilsonLoops}. The entropies provide novel and distinctive features in the presence of the Wilson loops. 
\\ \vspace{-0.1in}

{\bf Mutual Information:} 
Mutual (R\'enyi) information measures the entanglement between two intervals, $A$ and $B$ of length $\ell_A$ and $\ell_B$, separated by $\ell_C$. It is given by 
\begin{align}
I_n(A,B) = S_n (A) + S_n (B) - S_n (A\cup B) \;.
\end{align}
This is free of UV divergences and finite. 

Mutual information shares the same functional dependences on $\mu, J$ and $w$ as those of the entropies. For example, the spin structure dependent mutual information for the anti-periodic fermions is given (with $J\=\alpha\= w\=0$) in the low temperature limit \cite{Kim:2017ghc}.  
\begin{align}\label{MutualInfo}
I^\mu_{n\to 1} &(A,B) 
=\sum_{l=1}^{\infty} \frac{2 (-1)^{l-1}}{l} 
\frac{\cosh (l\beta\mu)}{\sinh ( \frac{l\beta}{2} )}  \\
&\times \bigg[n\-  \frac{\sin \left(\! \frac{l\ell_A}{2L}\right)}{\sin \left(\! \frac{l}{n} \frac{\ell_A}{2L}\right)} \-  \frac{\sin \left(\! \frac{l\ell_B}{2L}\right)}{\sin \left(\! \frac{l}{n} \frac{\ell_B}{2L}\right)} +  \frac{\sin \big(\! \frac{l(\ell_A+\ell_B)}{2L}\big)}{\sin \left(\! \frac{l}{n} \frac{\ell_A+\ell_B}{2L}\right)}\! \bigg]. \nonumber
\end{align} 
Note that the first line is the same as that of \eqref{SAmuNSRenyi} demonstrating the same functional dependence of $\mu$. Surprisingly, the spin structure dependent mutual information is independent of the separation between the two intervals. These results are also valid in the presence of the current source and Wilson loops. 
 
{\it Acknowledgements:} 
We thank John Cardy, Paul de Lange, Mitsutoshi Fujita, Elias Kiritsis, and Alfred Shapere  
for helpful discussions. We are especially thankful to Sumit Das for numerous discussions and valuable comments on the draft.
This work is partially supported by NSF Grant PHY-1214341.

\bigskip

\onecolumngrid \vspace{-0.2in}
\section*{Appendix}

\subsection{Partition function \& Correlation function} \label{sec:PartitionFunction}

We construct the partition function of 1+1 dimensional Dirac fermions in the presence of current source $J$ and chemical potential $\mu$ based on previous studies \cite{DiFrancesco:1997nk}\cite{Hori:2003ic}. In particular, we use the equivalence between the twisted boundary conditions and the background gauge fields to build up the partition function for the fermions. 

Consider the action for Dirac field $\psi$   
\begin{align} \tag{S1}
\mathcal S = \frac{1}{2\pi} \int d^2 x ~i  \bar \psi  \gamma^\mu \left(\partial_\mu + i A_\mu \right) \psi \;, 
\end{align}
where $\mu=0, 1$ are the time and space coordinates with $\gamma^0 = \sigma_1, \gamma^1 = -i \sigma_2$ in terms of Pauli matrices 
\[ \gamma^0 = \left( \begin{array}{cc}
~0~ & ~1~ \\
1 & 0 \end{array} \right)\;, \qquad 
\gamma^1 = \left( \begin{array}{cc}
~0~ & -1~ \\
1 & 0 \end{array} \right)\;,
\] 
$\bar \psi = \psi^\dagger \gamma^0$, 
and constant background gauge fields $A_0 = \mu, ~A_1 = J $ that are identified as chemical potential and current source. 

To compute the partition function, we consider a torus with the modular parameter $\tau = \tau_1 + i \tau_2$. Thus the space of coordinates $\zeta = \frac{1}{2\pi} (s + i t)$ is identified as $\zeta \equiv \zeta + 1 \equiv \zeta + \tau$. Now $s$ is a spatial coordinate with a circumference $2\pi L$ (with $L=1$ in this appendix), while $t$ is the Euclidean time with periodicity $2\pi \tau_2 = \beta = 1/T$. We decompose the Dirac field 
\[ \psi  = \left( \begin{array}{c}
~\psi_-~ \\
\psi_+ \end{array} \right)\;,
\] 
and consider twisted boundary conditions 
\begin{align}\label{TwistedeBC}\tag{S2}
\psi_\pm (t,s) &= e^{-2\pi i a} \psi_\pm (t,s+2\pi) 
= e^{-2\pi i b} \psi_\pm (t+2\pi \tau_2, s+2\pi \tau_1)  \;. 
\end{align}
There exists an equivalent description. It has the periodic Dirac fermion with the following flat gauge connection $\tilde A_\mu$ on a torus   
\begin{align}\label{FlatConnection}\tag{S3}
\tilde A = \tilde A_\mu dx^\mu = a ds + \frac{b -a\tau_1}{\tau_2} dt \;. 
\end{align}
Thus, in this equivalent description, one can identify the chemical potential and current source 
\begin{align} \tag{S4}
a = \tilde J \;,  \qquad b = \tau_1 \tilde J + i \tau_2 \tilde \mu \;.
\end{align}

Partition function is a trace of the Hilbert space constructed with the twisted periodic boundary condition for $s \!\sim\! s\+ 2\pi$ along with the Euclidean time evolution $t \!\to\! t\+2\pi \tau_2$ represented by the operator $ e^{-2\pi \tau_2 H}$, where $H$ is Hamiltonian. The latter also induces the space translation $s \!\to\! s \- 2\pi \tau_1$ represented by the operator $ e^{-2\pi i \tau_1 P}$ (with momentum operator $P $) together with a phase rotation due to the presence of fermion $ e^{-2\pi i (b-1/2) F_A}$, where $F_A \= \frac{1}{2\pi} \int ds ( \psi_+^\dagger \psi_+ \- \psi_-^\dagger \psi_-)$ is a fermion number. The partition function with the twisted parameters $a$ and $b$ is $Z_{[a,b]} = Tr \big[e^{-2\pi i (b-1/2) F_A} e^{-2\pi i \tau_1 P} e^{-2\pi \tau_2 H} \big]$. This can be written as 
\begin{align} \tag{S5}
Z_{[a,b]} = \left|\eta(\tau) \right|^{-2} \big|\vartheta \genfrac[]{0pt}{1}{1/2-a}{b-1/2 } (0,\tau)\big|^2
\;,
\end{align}   
where Dedekind function $\eta(\tau) = q^{\frac{1}{24}} \prod_{n=1}^\infty (1 - q^n)$, Jacobi function $\vartheta \genfrac[]{0pt}{1}{\alpha}{\beta} (z|\tau) = \sum_{n \in \mathbb Z} q^{\frac{(n + \alpha)^2}{2}} e^{2\pi i (z+\beta)(n + \alpha)}$, and $ q= e^{2\pi i \tau}$.

Focusing on the anti-periodic boundary conditions on both circles, $a=1/2, b=1/2$, the partition function is 
\begin{align} \tag{S6}
Z_{[\frac{1}{2},\frac{1}{2}]} = \left|\eta(\tau) \right|^{-2} \Big|\vartheta_3 (z|\tau)\Big|^2 = \left|\eta(\tau) \right|^{-2} \Big|\sum_{n \in \mathbb Z} q^{\frac{n^2}{2}} \Big|^2
\;. 
\end{align}
The following notations are also used in the literature. $\vartheta_3(z|\tau) = \vartheta \left[\substack{0 \\ 0} \right] (z|\tau)$, $\vartheta_2(z|\tau) = \vartheta \left[\substack{1/2 \\ 0} \right] (z|\tau)$, $\vartheta_4(z|\tau) = \vartheta \left[\substack{0 \\ 1/2} \right] (z|\tau)$, and $\vartheta_1(z|\tau) = \vartheta \left[\substack{1/2 \\ 1/2} \right] (z|\tau) $. $\vartheta_2(z|\tau)$ is related to the periodic spatial circle and anti-periodic time circle, while $\vartheta_3(z|\tau)$ to both the anti-periodic spatial and time circles. We focus on $\vartheta_2$ and $\vartheta_3$ because physical fermions are described by the anti-periodic temporal circle.

In the presence of current source $J$ and chemical potential $\mu$, the partition function has a more general form 
\begin{align} \tag{S7}
Z_{[a,b]}^{\mu, J} = Tr \left[e^{2\pi i (\tau_1 J + i \tau_2 \mu + b-\frac{1}{2}) F_A} e^{\-2\pi i \tau_1 P} e^{\-2\pi \tau_2 H} \right] 
\;.
\end{align}   
This can be understood by examining the Dirac action in the presence of the background gauge fields as well as the effective flat gauge connection $\tilde A$ given in \eqref{FlatConnection} 
\begin{align} \tag{S8}
\tilde {\mathcal S} = \frac{1}{2\pi} \int d^2 x ~i  \bar \psi  \gamma^\mu \left(\partial_\mu + i A_\mu + i \tilde A_\mu\right) \psi \;, 
\end{align}
where the Dirac field has the periodic boundary condition.  
In particular, the mode expansion of the fermion field depends not only on the twisted boundary condition, but also on the presence of current source $J$. For example, 
\begin{align} \tag{S9}
\psi_- = \sum_{r \in \mathbb Z + a} \psi_r (t) e^{irs} \quad \to \quad \sum_{\tilde r \in \mathbb Z + a + J} \psi_r (t) e^{i \tilde r s} \;.
\end{align}
Taking this into account, the partition function is
\begin{align} \tag{S10}
Z_{[a,b]}^{\mu, J} 
&=\left|\eta(\tau) \right|^{-2} \big|\vartheta \genfrac[]{0pt}{1}{1/2-a- J}{b-1/2 } (\tau_1 J +i\tau_2 \mu|\tau)\big|^2
\;.
\end{align} 
Note that the current source has two distinct contributions. One of them is through the modification of the Hilbert space. 
The periodicity of temporal direction and the corresponding thermal boundary condition are not modified. With this we can construct the two point correlation functions in the presence of chemical potential $\mu$ and current source $J$ following \cite{DiFrancesco:1997nk}. The correlation functions have been explicitly written in \cite{Kim:2017ghc}. 

At this point, we pause to think about constructing $n$-replica copy of the fermions in the presence of the Wilson loops. First, we consider the $n$ fermions without the Wilson loop. When the $m$-th fermion $\tilde \psi_m$ ($m=1,2,\cdots, n$) crosses the branch cut connecting two points $u$ and $v$ in 2 dimensional Riemann space, we have the following identifications 
\begin{align} \tag{S11}
\tilde \psi_m(e^{2\pi i} (x-u)) = \tilde \psi_{m+1} (x-u) \;, \qquad \tilde \psi_m(e^{2\pi i} (x-v)) = \tilde \psi_{m-1} (x-v) \;,
\end{align}	  
where $x$ is a complexified coordinate. 
These boundary conditions can be diagonalized by defining $n$ new fields  
\begin{align} \tag{S12}
\psi_k = \frac{1}{n} \sum_{m=1}^n e^{2\pi i m k } \tilde \psi_m  \;.
\end{align}	 
For the new field, the boundary condition becomes 
\begin{align} \tag{S13}
\psi_k(e^{2\pi i} (x-u)) = e^{2\pi i k/n} \psi_{k} (x-u) \;, \qquad \psi_k(e^{2\pi i} (x-v)) = e^{-2\pi i k/n} \psi_{k} (x-v) \;,
\end{align}	 
where $ k=-(n-1)/2, -(n-3)/2, \cdots, (n-1)/2$. The phase shift $e^{2\pi i k/n}$ is generated by the standard twist operator $\sigma_{k/n}$ with a conformal dimension $ \frac{1}{2} \frac{k^2}{n^2}$. The full twist operator is $\sigma_n = \Pi  \sigma_{k/n}$. 

In \cite{Belin:2013uta}, the authors notice that the boundary condition can be further generalized to include a global phase rotation $\tilde \psi \to e^{i w} \tilde \psi$. This phase can be added to the boundary condition for $\tilde \psi$ as 
\begin{align} \tag{S14}
\tilde \psi_m(e^{2\pi i} (x-u)) = e^{i w} \tilde \psi_{m+1} (x-u) \;, \qquad \tilde \psi_m(e^{2\pi i} (x-v)) = e^{-i w} \tilde \psi_{m-1} (x-v) \;.
\end{align}	  
This additional phase is added uniformly, the diagonal fields has the same effect. 
\begin{align}\label{GeneralTWB} \tag{S15}
\psi_k(e^{2\pi i} (x-u)) = e^{2\pi i k/n+iw} \psi_{k} (x-u) \;, \qquad \psi_k(e^{2\pi i} (x-v)) = e^{-2\pi i k/n-iw} \psi_{k} (x-v) \;.
\end{align}	 

These phase shifts can be handled by the generalized twist operators $\sigma_{w,k}$ with an electric parameter $w$ and a magnetic parameter $k$. Their name for the electromagnetic operators come from \cite{DiFrancesco:1997nk} along with the detailed expressions for the correlation functions. The expression for the two point functions of the electromagnetic operators $ \sigma_{w,k} (u)$ and $\sigma_{-w,-k} (v)$ with an electric charge $\frac{w}{2\pi} + l_k$ and a magnetic charge $\frac{k}{n}$, in the presence of the current source $J$ and the chemical potential $\mu$ as well as the twisted boundary conditions $a, b$, are given by  
\begin{align} \label{TopologicalCorrelatorTwistedOperators} \tag{S16}
\langle \sigma_{w,k} (u) \sigma_{-w,-k} (v) \rangle_{a,b,J,\mu} = \Big| \frac{2\pi \eta (\tau)^3 }{\vartheta [\substack{1/2 \\ 1/2 }](\frac{u-v}{2\pi L}|\tau)} \Big|
^{2\alpha_{w,k}^2}~ 
\Big| \frac{\vartheta [\substack{1/2-a-J \\ b-1/2 }](\frac{u-v}{2\pi L} \alpha_{w,k}+ \tau_1 J + i \tau_2 \mu|\tau)}{\vartheta [\substack{1/2-a-J \\ b-1/2 }](\tau_1 J + i \tau_2 \mu|\tau)} \Big|^2 \;. 
\end{align}
where $\alpha_{w,k} = \frac{k}{n} + \frac{w}{2\pi} + l_k$. The magnetic parameter $\frac{k}{n}$ comes into a play when we use the replica trick for $n$ copies of fermions in a single Riemann surface instead of a fermion on $n$ copies of Riemann surfaces \cite{Casini:2009sr}. The corresponding conformal dimension is given by 
\begin{align} \label{ConformalDimension} \tag{S17}
\Delta_{w,k}= \text{conformal dimension} = \frac{1}{2}  \alpha_{w,k}^2= \frac{1}{2} \left( \frac{k}{n} + \frac{w}{2\pi} + l_k \right)^2  \;.
\end{align}	 
The constant $l_k$ stems from the fact that the boundary condition \eqref{GeneralTWB} has an intrinsic ambiguity. This constant is used to minimize the conformal dimension of the twist operator such that $ -\frac{1}{2} \leq \alpha_{w,k} \leq \frac{1}{2}$. This is closely related to the topological transitions discussed in the main body. It is interesting to mention that the dependence of the background gauge fields in entanglement entropy can be systematically introduced by using the twisted boundary conditions represented by the parameters $a$ and $b$ \cite{Kim:2017ghc}\cite{Kim:toAppear} .

\subsection{Example Computations for anti-periodic fermions in the zero temperature limit} \label{sec:AntiPeriodicZeroEE}

We derive the equation \eqref{SAmuNS} in the main body. Entanglement entropy for the anti-periodic fermions is given by 
\begin{align}\tag{S18}
S_{n\to 1}^\mu &= \lim_{n\to 1} \Big(  \frac{1}{1- n} \bigg[ \sum_{k=-\frac{n-1}{2}}^{\frac{n-1}{2}} \log \Big|\frac{\vartheta_3 (\frac{k}{n} \frac{\ell_t}{L}+\frac{i\beta\mu}{2\pi} |i\beta )}{\vartheta_3 (\frac{i\beta\mu}{2\pi} |i\beta)} \Big|^2 \bigg] \Big) \;, 
\end{align} 
where $\beta = 2\pi \tau_2$ and $\ell_t = u-v $. 
Using the product representation 
$
\vartheta_3 (z|\tau) = \prod_{m=1}^\infty (1 - q^m)(1 + y q^{m-1/2})(1 + y^{-1} q^{m-1/2}),
$
with $y_1 = e^{- \beta\mu + 2\pi i \frac{k}{n} \frac{\ell_t}{L}}, ~y_2 = e^{- \beta\mu}, ~ q=e^{-\beta}$. We compute the R\'enyi entropy at low temperature limit, $\beta \to \infty$, 
\begin{align}\label{app2}\tag{S19}
S_n^\mu &= \frac{1}{1- n} \bigg[ \sum_{k=-\frac{n-1}{2}}^{\frac{n-1}{2}} \log \Big|\prod_{m=1}^{\infty} \frac{(1 - q^m)(1 + y_1 q^{m-1/2})(1 + y_1^{-1} q^{m-1/2})}{(1 - q^m)(1 + y_2 q^{m-1/2})(1 + y_2^{-1} q^{m-1/2})} \Big|^2 \bigg]  \\
&= \frac{1}{1- n} \bigg[ \sum_{m=1}^{\infty} \sum_{k=-\frac{n-1}{2}}^{\frac{n-1}{2}} \sum_{l=1}^{\infty} \frac{(-1)^{l-1}}{l} \left( [y_1^l +y_1^{l*} - y_2^l - y_2^{l*}] q^{l(m-1/2)} + [y_1^{-l} + y_1^{-l*} - y_2^{-l}  - y_2^{-l*} ]q^{l(m-1/2)}  \right) \bigg]  \nonumber \\
&= \frac{2}{1- n} \bigg[ \sum_{l,m=1}^{\infty} \sum_{k=-\frac{n-1}{2}}^{\frac{n-1}{2}}  \frac{(-1)^{l-1}}{l} 
\left( e^{-l\beta\mu} + e^{l\beta\mu}\right) e^{- l\beta(m-1/2)} \left[ \cos \left(2\pi \frac{k}{n} \frac{\ell_t}{L}l\right) -1 \right]  \bigg]  \nonumber  \\
&= \frac{2}{1- n} \bigg[ \sum_{l,m=1}^{\infty} \frac{(-1)^{l-1}}{l} 
\left( e^{-l\beta\mu} + e^{l\beta\mu}\right) e^{- l\beta(m-1/2)} \left[-n +  \sin \left(\pi \frac{\ell_t}{L}l\right) \csc \left(\frac{\pi}{n} \frac{\ell_t}{L}l\right) \right]  \bigg]  \nonumber  
\end{align}
Entanglement entropy is given by 
\begin{align} \tag{S20}
S_{n\to 1}^\mu &=2\sum_{l=1}^{\infty} \frac{(-1)^{l-1}}{l} 
\frac{\cosh\left(l\beta\mu\right)}{\sinh\left(l\beta/2\right)} \left[1- \pi l \frac{\ell_t}{L} \cot \left(\pi l \frac{\ell_t}{L} \right) \right]  
\;. \nonumber 
\end{align} 

\subsection{Example Computations for anti-periodic fermions in the large radius limit} \label{sec:AntiPeriodicLargeEE}

In this chapter, we derive the equation \eqref{EELargeSpaceLimitNSSectorRenyi} of the letter. Using $\vartheta_3 (z|\tau)$ along with the identification $z_1 = i\frac{\beta\mu}{2\pi}, ~z_2 = i\frac{\beta\mu}{2\pi} + \frac{k}{n} \frac{\ell_t}{L}, ~ q=e^{-2\pi \beta}$.
For $\ell_t/L \ll 1$, one has  
\begin{align}\tag{S21}
&\cos (i \beta\mu + 2\pi \frac{k}{n} \frac{\ell_t}{L}) \= 
\cosh (\beta\mu) \- 2\pi i \frac{k}{n} \frac{\ell_t}{L} \sinh (\beta\mu)  \-\frac{1}{2} (2\pi \frac{k}{n} \frac{\ell_t}{L})^2 \cosh (\beta\mu) + \cdots   \;, \\
&1 \+ q^{2m\- 1} \+ 2 \cos (2\pi z_2) q^{m\-1/2} 
\=  
1 \+ q^{2m\- 1} \+ 2 \cos (2\pi z_1) q^{m\-1/2} \nonumber \\
&\hspace{1.85in} -2 q^{m\-1/2} \left(2\pi i \frac{k}{n} \frac{\ell_t}{L} \sinh ( \beta\mu)  \+\frac{1}{2} (2\pi \frac{k}{n} \frac{\ell_t}{L})^2 \cosh ( \beta\mu)\right) \+ \cdots \;. \nonumber 
\end{align}

Then the R\'enyi entropy has the form 
\begin{align}\label{app4}\tag{S22}
S^\mu_n 
&= \frac{1}{1- n} \bigg[ \sum_{k=-\frac{n-1}{2}}^{\frac{n-1}{2}} \log \Big|\frac{\vartheta_3 (\frac{k}{n} \frac{\ell_t}{L}+\frac{i\beta\mu}{2\pi} |i\beta )}{\vartheta_3 (\frac{i\beta\mu}{2\pi} |i\beta)} \Big|^2 \bigg]    \nonumber \\
&=  \frac{1}{1- n} \bigg[  \sum_{m=1}^{\infty} \sum_{k=-\frac{n-1}{2}}^{\frac{n-1}{2}} \log \bigg| 1 - \frac{  2\pi i \frac{k}{n} \frac{\ell_t}{L} \sinh (\beta\mu) + 2\pi^2 (\frac{k}{n} \frac{\ell_t}{L})^2 \cosh (\beta\mu)}{\cosh (\beta [m-1/2]) + \cosh (\beta \mu) } + \cdots \bigg|^2 \bigg]  \nonumber \\
&=  \frac{-1}{1- n} \bigg[ \sum_{m=1}^{\infty} \sum_{k=-\frac{n-1}{2}}^{\frac{n-1}{2}}  (\pi\frac{k}{n} \frac{\ell_t}{L})^2   \Big[ \frac{ 4 \cosh (\beta\mu)}{\cosh (\beta [m-1/2]) + \cosh (\beta \mu) } \!+\!  \frac{ 4 \sinh^2 (\beta\mu)}{(\cosh (\beta [m-1/2]) \!+\! \cosh (\beta \mu) )^2 }\Big] \bigg]   \!+\! \cdots\nonumber  \\ 
&= \frac{(n+1)\pi^2}{12 n} \frac{\ell_t^2}{L^2} \sum_{m=1}^{\infty} \frac{ 4 + 4 \cosh (\beta\mu) \cosh (\beta [m-1/2])}{(\cosh (\beta [m-1/2]) + \cosh (\beta \mu) )^2 } + \mathcal O \left(\frac{\ell_t}{L} \right)^4 \;, \nonumber 
\end{align} 
where we use the condition $\frac{\ell_t}{L} \ll 1 $ for a general temperature. In the third line, we use the fact 
\begin{align} \label{app5}\tag{S23}
\log \Big| 1 + i a \frac{\ell_t}{2\pi L} + b \frac{\ell_t^2}{4\pi^2 L^2} + \mathcal O (\frac{\ell_t^3}{8\pi^3 L^3} ) \Big|^2 
= (a^2  + 2b) \frac{\ell_t^2}{4\pi^2 L^2} + \mathcal O (\frac{\ell_t^4}{16 \pi^4 L^4} ) \;. 
\end{align} 
Similar results happen for all the spin dependent entropies in the presence of chemical potential, current source or the Wilson loop parameter. 


\begin{thebibliography}{99}

\bibitem{Einstein:1935rr} 
  A.~Einstein, B.~Podolsky and N.~Rosen,
  ``Can quantum mechanical description of physical reality be considered complete?,''
  Phys.\ Rev.\  {\bf 47}, 777 (1935).
  doi:10.1103/PhysRev.47.777

\bibitem{Horodecki:2009zz} 
  R.~Horodecki, P.~Horodecki, M.~Horodecki and K.~Horodecki,
  ``Quantum entanglement,''
  Rev.\ Mod.\ Phys.\  {\bf 81}, 865 (2009)
  doi:10.1103/RevModPhys.81.865
  [quant-ph/0702225].

\bibitem{Calabrese:2004eu} 
  P.~Calabrese and J.~L.~Cardy,
  ``Entanglement entropy and quantum field theory,''
  J.\ Stat.\ Mech.\  {\bf 0406}, P06002 (2004)
  doi:10.1088/1742-5468/2004/06/P06002
  [hep-th/0405152].

\bibitem{Ryu:2006bv} 
  S.~Ryu and T.~Takayanagi,
  ``Holographic derivation of entanglement entropy from AdS/CFT,''
  Phys.\ Rev.\ Lett.\  {\bf 96}, 181602 (2006)
  doi:10.1103/PhysRevLett.96.181602
  [hep-th/0603001].
    
\bibitem{Rangamani:2016dms} 
  M.~Rangamani and T.~Takayanagi,
  ``Holographic Entanglement Entropy,''
  Lect.\ Notes Phys.\  {\bf 931}, 2017
  doi:10.1007/978-3-319-52573-0
  [arXiv:1609.01287 [hep-th]].
  
\bibitem{QunatumInformation} 
  M. A. Nielsen and I. L. Chuang,
  ``Quantum Computation and Quantum Information,''
  Cambridge Univ. Press, 2010

\bibitem{Renyi} 
A. Renyi, ``On measures of entropy and information,'' in
Proc. Fourth Berkeley Symp. Math. Stat. Prob., 1960,
Vol. 1, 547, Berkeley, 1961. University of California Press.

\bibitem{Vidal:2002rm} 
  G.~Vidal, J.~I.~Latorre, E.~Rico and A.~Kitaev,
  ``Entanglement in quantum critical phenomena,''
  Phys.\ Rev.\ Lett.\  {\bf 90}, 227902 (2003)
  doi:10.1103/PhysRevLett.90.227902
  [quant-ph/0211074].

\bibitem{Kitaev:2005dm} 
  A.~Kitaev and J.~Preskill,
  ``Topological entanglement entropy,''
  Phys.\ Rev.\ Lett.\  {\bf 96}, 110404 (2006)
  doi:10.1103/PhysRevLett.96.110404
  [hep-th/0510092].

\bibitem{Levin:2006zz} 
  M.~Levin and X.~G.~Wen,
  ``Detecting Topological Order in a Ground State Wave Function,''
  Phys.\ Rev.\ Lett.\  {\bf 96}, 110405 (2006).
  doi:10.1103/PhysRevLett.96.110405

\bibitem{Casini:2012ei} 
  H.~Casini and M.~Huerta,
  ``On the RG running of the entanglement entropy of a circle,''
  Phys.\ Rev.\ D {\bf 85}, 125016 (2012)
  doi:10.1103/PhysRevD.85.125016
  [arXiv:1202.5650 [hep-th]].

\bibitem{Experiment}
 R. Islam,	R. Ma,	P. M. Preiss, M. E. Tai,	A. Lukin, M. Rispoli	and M. Greiner, 
``Measuring entanglement entropy in a quantum many-body system,''
Nature, {\bf 528}, 77 (2015).

\bibitem{Holzhey:1994we} 
C.~Holzhey, F.~Larsen and F.~Wilczek,
``Geometric and renormalized entropy in conformal field theory,''
Nucl.\ Phys.\ B {\bf 424}, 443 (1994)
doi:10.1016/0550-3213(94)90402-2
[hep-th/9403108].

\bibitem{Bennett:1995tk} 
C.~H.~Bennett, H.~J.~Bernstein, S.~Popescu and B.~Schumacher,
``Concentrating partial entanglement by local operations,''
Phys.\ Rev.\ A {\bf 53}, 2046 (1996)
doi:10.1103/PhysRevA.53.2046
[quant-ph/9511030].

\bibitem{Calabrese:2009qy} 
P.~Calabrese and J.~Cardy,
``Entanglement entropy and conformal field theory,''
J.\ Phys.\ A {\bf 42}, 504005 (2009)
doi:10.1088/1751-8113/42/50/504005
[arXiv:0905.4013 [cond-mat.stat-mech]].

\bibitem{Casini:2009sr} 
H.~Casini and M.~Huerta,
``Entanglement entropy in free quantum field theory,''
J.\ Phys.\ A {\bf 42}, 504007 (2009)
doi:10.1088/1751-8113/42/50/504007
[arXiv:0905.2562 [hep-th]].

\bibitem{Ogawa:2011bz} 
  N.~Ogawa, T.~Takayanagi and T.~Ugajin,
  ``Holographic Fermi Surfaces and Entanglement Entropy,''
  JHEP {\bf 1201}, 125 (2012)
  doi:10.1007/JHEP01(2012)125
  [arXiv:1111.1023 [hep-th]].
  
\bibitem{Herzog:2013py} 
  C.~P.~Herzog and T.~Nishioka,
  ``Entanglement Entropy of a Massive Fermion on a Torus,''
  JHEP {\bf 1303}, 077 (2013)
  doi:10.1007/JHEP03(2013)077
  [arXiv:1301.0336 [hep-th]].

\bibitem{CardySlides:2016} 
  J.~Cardy,
  ``Entanglement in CFTs at Finite Chemical Potential,'' \href{http://www2.yukawa.kyoto-u.ac.jp/~entangle2016/YCardy.pdf}{presentation} at the Yukawa International Seminar 2016 (YKIS2016) ``Quantum Matter, Spacetime and Information.'' 

\bibitem{Belin:2013uta} 
A.~Belin, L.~Y.~Hung, A.~Maloney, S.~Matsuura, R.~C.~Myers and T.~Sierens,
``Holographic Charged Renyi Entropies,''
JHEP {\bf 1312}, 059 (2013)
doi:10.1007/JHEP12(2013)059
[arXiv:1310.4180 [hep-th]].

\bibitem{Arias:2014ksa} 
  R.~E.~Arias, D.~D.~Blanco and H.~Casini,
  ``Entanglement entropy as a witness of the Aharonov-Bohm effect in QFT,''
  J.\ Phys.\ A {\bf 48}, no. 14, 145401 (2015)
  doi:10.1088/1751-8113/48/14/145401
  [arXiv:1409.3269 [hep-th]].

\bibitem{DiFrancesco:1997nk} 
  P.~Di Francesco, P.~Mathieu and D.~Senechal,
  ``Conformal Field Theory,'' 1997, Springer
  doi:10.1007/978-1-4612-2256-9

\bibitem{Hori:2003ic} 
  K.~Hori, S.~Katz, A.~Klemm, R.~Pandharipande, R.~Thomas, C.~Vafa, R.~Vakil and E.~Zaslow,
  ``Mirror symmetry,'' 2003
  American Mathematical Society

\bibitem{Kim:2017ghc} 
B.~S.~Kim,
``Entanglement Entropy with Background Gauge Fields,''
JHEP {\bf 1708}, 041 (2017)
doi:10.1007/JHEP08(2017)041
[arXiv:1706.07110 [hep-th]].

\bibitem{Kim:toAppear} 
B.~S.~Kim,
``Entanglement Entropy and Wilson loop,'' to appear. 

\bibitem{Casini:2005rm} 
H.~Casini, C.~D.~Fosco and M.~Huerta,
``Entanglement and alpha entropies for a massive Dirac field in two dimensions,''
J.\ Stat.\ Mech.\  {\bf 0507}, P07007 (2005)
doi:10.1088/1742-5468/2005/07/P07007
[cond-mat/0505563].

\bibitem{Nishioka:2009zj} 
T.~Nishioka, S.~Ryu and T.~Takayanagi,
``Holographic Superconductor/Insulator Transition at Zero Temperature,''
JHEP {\bf 1003}, 131 (2010)
doi:10.1007/JHEP03(2010)131
[arXiv:0911.0962 [hep-th]].

\end{thebibliography}
\end{document}